\documentclass[journal]{IEEEtran}
\usepackage{amsmath,amssymb,amsfonts}
\usepackage{booktabs}
\usepackage{graphicx}
\usepackage{xcolor}
\usepackage[hidelinks]{hyperref}
\usepackage{tikz}
\usetikzlibrary{arrows.meta,positioning,fit,backgrounds}
\usepackage{algorithm}
\usepackage{algpseudocode}
\usepackage{amsthm}
\newtheorem{theorem}{Theorem}
\newtheorem{definition}{Definition}

\newcommand{\actnone}{\textsf{act=none}}
\newcommand{\actyes}{\textsf{act}}
\newcommand{\tma}{TMA-NM}
\usepackage{pifont}
\newcommand{\cmark}{\ensuremath{\checkmark}}   % extracts as U+2713
\newcommand{\xmark}{\ensuremath{\times}}        % extracts as U+00D7
\newcommand{\pmark}{\ensuremath{\sim}}          % partial; legended in the caption
\newcommand{\orcidicon}[1]{%
    \href{https://orcid.org/#1}{%
        \includegraphics[width=10pt]{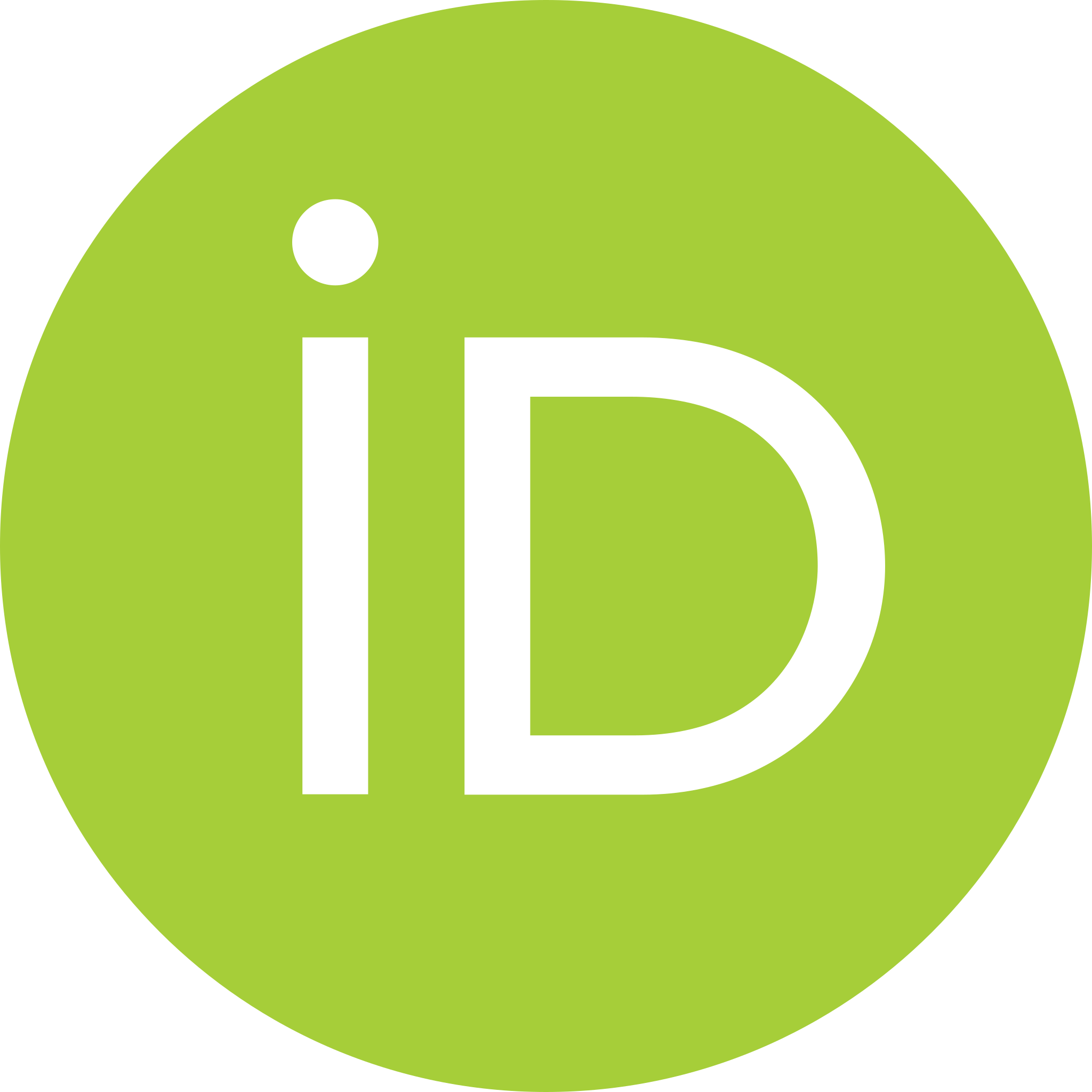}%
    }%
}

\begin{document}
\title{Securing LLM-Agent Long-Term Memory Against Poisoning:\\ Non-Malleable, Origin-Bound Authority with Machine-Checked Guarantees}

\author{
\IEEEauthorblockN{Yedidel Louck \orcidicon{0009-0008-5836-8736}}
\thanks{Yedidel Louck is with Department of Computer and Software Engineering, Ariel Cyber Innovation Center, Ariel University, Israel.\\
yedidel.louck@msmail.ariel.ac.il}
}

% \markboth{IEEE Transactions on Dependable and Secure Computing}%
% {Louck: Non-Malleable Memory Authority for LLM Agents}

\maketitle

\begin{abstract}
LLM agents increasingly rely on persistent long-term memory, which creates a critical
vulnerability that we study here: \emph{memory poisoning}. An adversary can store untrusted
content in one session that later steers a consequential action, such as a payment, a
setting change, or data exfiltration, in a future session. Existing defenses base a memory
item's authority to act on either its \emph{content} (detection or trust-scoring) or its
\emph{derivation history} (lineage). We show that both signals are \emph{malleable}. An
attacker can \emph{launder} an untrusted origin through three channels specific to LLM
agents: the agent's own summarization, a trusted-tool echo, and manufactured
corroboration. Each makes the content look benign and breaks or flips its derivation edge
to ``trusted.'' We formalize malleability for the memory write-retrieve-act pipeline and
prove a machine-checked separation theorem. No content- or
lineage-based defense is sound under laundering (T1), write-time origin binding is
necessary (T2), and non-malleable origin-bound authority with Sybil-resistant
corroboration-gated elevation is sufficient (T3). Our construction, \tma{} (Tamper-evident
Memory Authority, Non-Malleable), instantiates non-malleable information-flow control (IFC)
for LLM-agent memory. A cross-defense, cross-attack, and cross-model benchmark over eight
frontier models shows that existing defenses fail exactly where the theory predicts (up to 68\% laundering attack-success), while \tma{} reaches $0\%$ attack success on both
direct and laundering attacks across all models and channels, at full legitimate utility.
We release the benchmark, harness, and machine-checked TLA\textsuperscript{+} models to support reproducibility.
\end{abstract}

\begin{IEEEkeywords}
LLM agents, long-term memory, memory poisoning, information-flow control, capability
security, prompt injection, formal verification, benchmark.
\end{IEEEkeywords}

\section{Introduction}
\IEEEPARstart{P}{ersistent} memory is now a default capability of deployed LLM agents.
Systems such as MemGPT~\cite{packer2024memgptllmsoperatingsystems},
Mem0~\cite{chhikara2025mem0buildingproductionreadyai}, Generative
Agents~\cite{park2023generativeagentsinteractivesimulacra}, and
MemMachine~\cite{wang2026memmachinegroundtruthpreservingmemorypersonalized} let an
agent write facts, preferences, and instructions in one session and recall them in
later ones. This statefulness is also a vulnerability. A growing body of work
demonstrates \emph{memory poisoning}: sleeper/dormant poisoning that re-emerges across
unrelated sessions~\cite{pulipaka2026hiddenmemorysleepermemory} (echoing the
training-time sleeper-agent threat~\cite{hubinger2024sleeperagentstrainingdeceptive}),
memory control-flow
attacks that hijack tool selection~\cite{xu2026storagesteeringmemorycontrol},
retrieval corruption~\cite{zou2024poisonedragknowledgecorruptionattacks}, memory
injection~\cite{dong2026memoryinjectionattacksllm}, and agent
backdoors~\cite{chen2024agentpoisonredteamingllmagents,yang2024watchagentsinvestigatingbackdoor},
with reported injection success up to ${\sim}99.8\%$.

The defensive response has been limited~\cite{chu2026systematicsurveysecuritythreats}
and, we argue, aimed at the wrong target. Detection asks whether a memory ``looks
malicious,'' but the attacks show why this is hard. A poisoned memory is benign until it
activates, so there is nothing to detect at write or rest time, and an adaptive adversary
can disguise the trigger to survive content inspection at use
time~\cite{lin2026surveylongtermmemorysecurity}. The closest structural defense,
MemLineage~\cite{ouyang2026memlineagelineageguidedenforcementllm}, tracks provenance and
gates sensitive actions, but it re-derives its verdict at dispatch and offers no path for
legitimate untrusted information to ever act.

As a concrete example, suppose an agent reads a web page while researching a task and saves
a note for later. The page says to send the customer list to a given address whenever
backups are mentioned. The agent rewrites the page in its own words, and the note now reads
like the agent's own benign memory. Weeks later, in an unrelated session, the user asks
about backups, the agent recalls the note, and it exfiltrates the data. The untrusted web
origin has been \emph{laundered} by the agent's own summarization, so a defense that
inspects the note's content, or its now-dropped derivation edge, sees nothing wrong. This
is exactly the gap we close.

Our starting observation is that existing defenses decide a memory's authority to act from
a signal the adversary can \emph{transform}, namely its \emph{content} (detection or
trust-scoring) or a \emph{derivation edge} (lineage). We call such signals \emph{malleable},
and we call \emph{laundering} the act of erasing an item's untrusted origin through a
behavior-preserving transformation. An LLM-agent adversary has three such transformations.
In \emph{self-summarization} (L-a) the agent paraphrases the poison into its own note, so
the content looks benign and the derivation edge is dropped. In \emph{trusted-tool echo}
(L-b) a trusted tool returns attacker-controlled content. In \emph{manufactured
corroboration} (L-c) the adversary plants several untrusted items to fake a consensus.
Each lowers an item's apparent untrusted-ness without any genuine trusted endorsement.

We argue that authority to act must instead be \emph{non-malleable}, bound to a memory
item's true origin, which the trusted monitor records and no adversary transformation can
lower. Untrusted-origin content is non-actionable, and, crucially, this label
\emph{propagates}. An agent note derived from untrusted content, or a trusted-tool output
echoing it, inherits untrusted authority, which closes L-a and L-b. Authority rises only
through \emph{Sybil-resistant corroboration-gated elevation}, that is, through at least two
\emph{independent} trusted principals (for example an internal registry and a bank
confirmation, not two copies of the same web page) or a fresh action-bound user
authorization, never through content asserting its own legitimacy, which closes L-c. This is
a direct instantiation of nonmalleable information-flow control~\cite{cecchetti2017nonmalleable}
(IFC, the classical discipline that tracks how data of different trust levels may flow and
combine) for agent memory. It provides \emph{robust declassification}, so that an attacker who only
influences low-integrity data cannot cause a downgrade of authority, and \emph{transparent
endorsement}, so that integrity may be raised only by a principal that is itself trusted
on the endorsed data and never by the untrusted data vouching for itself.

We do not claim origin-tagging or cross-session enforcement themselves, since prior
systems already provide these (for example,
MemLineage~\cite{ouyang2026memlineagelineageguidedenforcementllm}). Our novelty is the
\emph{separation} (malleable defenses are unsound, non-malleable authority is sufficient),
the non-malleable construction that achieves it, and the benchmark that witnesses it,
detailed as C1--C5 below.

\textbf{Contributions.}
\begin{enumerate}
\renewcommand{\labelenumi}{C\arabic{enumi}.}
\item A formal model of LLM-agent memory authority + a definition of \emph{malleability}
via three LLM-specific laundering channels: self-summarization, trusted-tool echo,
manufactured corroboration (Section~\ref{sec:threat},\,Section~\ref{sec:formal}).
\item A \textbf{machine-checked separation theorem}: content- and malleable-lineage
defenses are unsound under laundering (T1), write-time origin binding is necessary within
the model (T2), and non-malleable origin-bound authority with Sybil-resistant
corroboration-gated elevation is sufficient (T3) (Section~\ref{sec:formal}).
\item A constructive \textbf{laundering attack} breaking all three malleable defense
classes (lineage, capability-IFC, and trust-scoring, instantiated by MemLineage,
CaMeL/Fides, and SuperLocalMemory), the empirical witness of T1 (Section~\ref{sec:eval}).
\item \textbf{MEM-INV-Bench}, the broadest cross-defense $\times$ attack $\times$ model
memory benchmark we are aware of, exhibiting the \emph{theory$\leftrightarrow$benchmark
correspondence} in which each defense class fails exactly where the theorem predicts
(Section~\ref{sec:eval}).
\item A released, reproducible artifact (benchmark, harness, machine-checked
proofs).\footnote{Our artifact is publicly available: code, harness, and
TLA\textsuperscript{+} models at \url{https://github.com/yedidel/mem-inv-bench}, and the
benchmark data at \url{https://huggingface.co/datasets/anonymos-2321135/MEM-INV-Bench}.
}
\end{enumerate}

\section{Threat Model}\label{sec:threat}
A stored memory item is a tuple $(\textit{content}, \textit{origin}, \textit{scope},
t_{\text{write}}, \textit{act\_class})$, where $\textit{origin}\in\{$user, trusted\_tool,
agent, untrusted\_external$\}$ and $\textit{act\_class}\in\{$none, inform, act$\}$. The
agent runs a \emph{write} $\rightarrow$ \emph{retrieve} $\rightarrow$ \emph{act} pipeline,
where an act is a response and/or a consequential action such as a tool call, payment,
setting change, or data egress.

The adversary controls \textsf{untrusted\_external} content that the agent may store, for
instance a document, a web page, a tool output, or another user's message. The adversary
cannot forge \textit{origin} or \textit{scope}, which the trusted monitor sets at write
time, cannot break cryptography, and does not control the user's authorization channel.

\begin{center}
\fbox{\begin{minipage}{0.94\columnwidth}\small
\textbf{Assumption A1 (origin-labeling oracle).} The monitor assigns each write its
true \textit{origin} from the \emph{channel} it arrives on (user prompt, authenticated
tool, agent self-write, untrusted ingestion), never from its content. We assume the
channel itself is authenticated, so labels are correct at write time. Concretely, a
\textsf{trusted} tool channel is realized by mutually authenticated transport (mTLS),
audience-bound OAuth tokens, or signed tool responses, so that origin derives from a
verified channel identity rather than from message text, and the monitor can flag
\emph{label drift} when a principal's claimed origin no longer matches its authenticated
identity. All of our
guarantees are \emph{conditional} on A1. If a trusted channel is itself compromised, its
outputs are mislabeled \textsf{trusted} and the binding inherits that error, which is
exactly why elevation requires \emph{two independent} trusted principals
(Section~\ref{sec:design}, M3) and so bounds the damage of any single compromised channel.
\end{minipage}}
\end{center}

We consider five attack classes: \emph{sleeper} (store now, trigger later),
\emph{control-flow} (a retrieved item redirects a tool or action), \emph{direct} (poison
that biases an answer), \emph{data-exfiltration} (a stored instruction causes egress), and
\emph{cross-agent} shared memory, which is out of scope and is the province of
agent-communication-protocol security~\cite{louck2025security,louck2025improving}. The
guarantee we target is that,
for any reachable state, no \textsf{untrusted\_external} item causes a consequential action
without independent trusted corroboration or a fresh user authorization bound to that
action. Non-consequential responses may still surface untrusted memory with provenance, an
honest scope boundary under which answer-biasing is \emph{mitigated, not eliminated}.
Write-time binding applies the Biba integrity principle (low-integrity inputs may not raise
an item's trust) in a temporal setting, combined with Denning-style lattice information
flow (trust levels form an ordered lattice along which data may move) and an
enforcement-monitor view of the act gate.

\section{The Non-Malleable Construction (\tma{})}\label{sec:design}
\begin{figure}[t]
\centering
\begin{tikzpicture}[
  font=\footnotesize,
  box/.style={draw,rounded corners,minimum height=7mm,inner sep=3pt,align=center},
  trust/.style={box,fill=green!25},
  untrust/.style={box,fill=red!25},
  mon/.style={draw,thick,rounded corners,fill=blue!15,align=center},
  >={Stealth[]}]
\node[untrust] (ext) {untrusted\\source};
\node[trust, below=4mm of ext] (tool) {trusted\\tool};
\node[mon, right=10mm of ext, minimum width=20mm, minimum height=20mm] (m)
  {\textbf{\tma{}}\\monitor};
\node[box, right=12mm of m, yshift=5mm] (mem) {long-term\\memory};
\node[box, right=12mm of m, yshift=-9mm] (gate) {act\\gate};
\node[box, right=10mm of gate] (act) {tool /\\payment};
\draw[->] (ext) -- node[above,sloped]{\actnone} (m.west|-ext);
\draw[->] (tool) -- node[below,sloped]{\actyes} (m.west|-tool);
\draw[->] (m.east|-mem) -- node[above]{write-bind} (mem);
\draw[->] (mem) to[bend left=15] node[right]{retrieve} (gate);
\draw[->] (gate) -- node[above]{verdict} (act);
\begin{scope}[on background layer]
\node[draw,dashed,rounded corners,fit=(m)(mem)(gate),inner sep=4mm,
  label=below:{append-only verdict log (M4)}] {};
\end{scope}
\end{tikzpicture}
\caption{\tma{} mediates write, retrieve, and act. Authority (\textit{act\_class}) is
bound at write time from origin (M1), and untrusted items are \actnone. The act gate
permits a consequential action only if its value is trusted-derived or elevated by
independent corroboration (M3), and authority propagates non-malleably (M2). All decisions
are logged (M4).}
\label{fig:arch}
\end{figure}
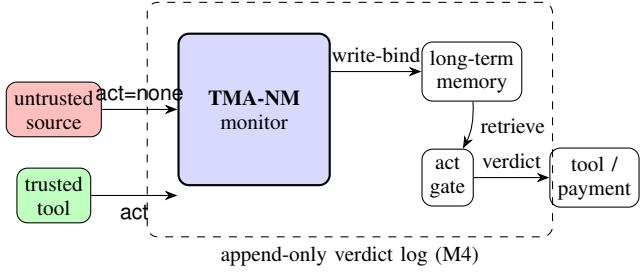

We call our construction \tma{}, for \emph{Tamper-evident Memory Authority,
Non-Malleable}: authority is bound to origin (\emph{Authority}), the binding cannot be
lowered by any laundering transformation (\emph{Non-Malleable}), and every verdict is
appended to a tamper-evident hash chain (\emph{Tamper-evident}). It realizes
the sufficiency direction: origin-bound authority whose label is non-malleable under the
three laundering channels, plus Sybil-resistant corroboration-gated elevation.
\emph{(Malleability is defined formally in Section~\ref{sec:formal}, and this section presents the
construction that resists it.)} Fig.~\ref{fig:arch} shows the monitor mediating the
pipeline.

The construction has four mechanisms, sketched in Fig.~\ref{fig:arch}. The first,
\emph{write-time origin binding} (M1), sets $\textit{act\_class}=f(\textit{origin})$ at the
moment of writing, so that untrusted maps to none, agent to inform, and trusted\_tool or
user to act. The monitor records \textit{origin}, and the adversary cannot forge it.

Non-malleable propagation (M2) is the keystone of the design, and it closes L-a and L-b.
Authority is bound irrevocably to \emph{origin} at write time and never to content or
derivation edges. Any item derived from untrusted sources, whether through agent
summarization (L-a) or a trusted tool fed untrusted inputs (L-b), automatically inherits
the \emph{maximum untrust} level and remains \actnone, no matter how benign the resulting
text appears. This is robust declassification and transparent
endorsement~\cite{cecchetti2017nonmalleable} instantiated for memory, where a
transformation cannot lower authority.

For opaque tools the monitor does \emph{not} infer dataflow inside a black box. It
propagates untrust at the \emph{call boundary}, so a tool invocation's output inherits the
maximum untrust of the arguments and retrieved memory items passed into that call, namely
its \emph{declared input bindings}, which the monitor mediates. A trusted tool fed
attacker-controlled content thus emits an \textsf{untrusted} result, closing L-b, without
any tool-internal provenance hook. A tool's output is treated as \textsf{trusted} only when
it is on an allow-list of origin-authoritative tools and all of its bound inputs are
trusted, and otherwise the boundary rule conservatively taints. This is the standard sound
over-approximation in which taint may be raised but never silently lowered. Tighter,
sub-value propagation through structured payloads is the value-level capability-token
extension (Section~\ref{sec:lim}).

The third mechanism is \emph{Sybil-resistant corroboration-gated elevation} (M3), which
closes L-c. A consequential action whose security-relevant value is driven by an untrusted
item is permitted only when that value is corroborated by ${\ge}2$ \emph{independent}
trusted principals, meaning distinct registries or tools rather than repeated or echoed
content, or when a fresh action-bound user authorization is presented and consumed.
Repeated untrusted items (manufactured corroboration, L-c) do not count, elevation is
monotone, and an item cannot corroborate itself. This is the legitimate path that
preserves utility.

The user authorization is bound to the \emph{concrete} action rather than the session. It
is a single-use token over the tuple $(\textit{tool},\textit{value }v,\textit{amount},
\textit{nonce},\textit{timestamp})$, and the act gate executes only when the dispatched
action matches the tuple and the token is unspent. This rules out replay, cross-action
substitution, and payee-swap or TOCTOU (time-of-check to time-of-use) between proposal and
execution, because re-binding to the exact $v$ is part of the verdict and changing $v$
after authorization voids it.

The fourth mechanism is a \emph{tamper-evident verdict log} (M4). Writes, elevations, and
allow/deny \emph{verdicts} are appended to a hash chain, and retroactive edits break the
chain. Each append is $O(1)$, a single hash over the previous head, and adds no model call,
so logging does not affect the act-gate latency. Storage grows linearly in the number of
verdicts, and periodic Merkle checkpointing with compaction of pre-checkpoint entries
bounds it in practice, keeping the scheme viable at high throughput.

Algorithm~\ref{alg:gate} states the act-time decision.
\begin{algorithm}[t]
\caption{Act-time authorization for a consequential action $a$ (allow iff its value is
not untrusted-derived, OR is licensed by ${\ge}2$ independent trusted principals, OR a
fresh user authorization is presented, otherwise deny).}\label{alg:gate}
\begin{algorithmic}[1]
\State $v \gets$ security-relevant value of $a$ (e.g.\ payee, setting, recipient)
\State $U \gets \{$retrieved untrusted items that push $v\}$
\If{$U = \emptyset$} \Return \textsc{allow} \Comment{$v$ not from untrusted memory}
\EndIf
\State $C \gets \{$\emph{independent} trusted principals that license $v\}$
\If{$|C| \ge 2$} elevate each $u\in U$; \Return \textsc{allow} \Comment{M3}
\EndIf
\If{fresh user auth bound to $a$} consume it; \Return \textsc{allow}
\EndIf
\State \Return \textsc{deny} \Comment{untrusted-derived value, uncorroborated}
\end{algorithmic}
\end{algorithm}

Each laundering channel fails for a structural reason. Self-summarization (L-a) fails
because the agent's note inherits the maximum untrust of its sources (M2), so paraphrasing
into benign-looking text grants no authority. Trusted-tool echo (L-b) fails because a tool
output whose inputs are untrusted is conditionally tainted (M2), so echoing attacker
content does not launder it. Manufactured corroboration (L-c) fails because elevation
counts only \emph{independent} trusted principals (M3), so repeated untrusted items never
reach the threshold. Sleeper and disguised triggers are subsumed, since authority is bound
to origin rather than to timing or phrasing. Trust is relocated to the origin-labeling
boundary and to the independence of corroborators, and a fully compromised trusted tool is
the main residual risk (Section~\ref{sec:lim}). Non-consequential answer-biasing is out of
scope and is surfaced with provenance.

\section{Formal Model and the Separation Theorem}\label{sec:formal}
Authority labels form the integrity order
$\textsf{untrusted}\sqsubset\textsf{agent}\sqsubset\textsf{trusted}$, and a \emph{defense}
$D$ maps each memory item to a label in this order. Let $\mathcal{A}$ be the three
laundering channels (self-summarization L-a, trusted-tool echo L-b, manufactured
corroboration L-c), the adversary transformations available \emph{without} genuine trusted
or user consent. Following nonmalleable
IFC~\cite{cecchetti2017nonmalleable}:

\begin{definition}[Malleability]\label{def:mall}
$D$ is \emph{malleable} if there exist a memory item $x$ and a transformation
$\tau\in\mathcal{A}^{\ast}$ with $D(\tau(x))\sqsupset D(x)$, that is, $\tau$ raises the
effective authority without consent. $D$ is \emph{non-malleable} if
$D(\tau(x))\sqsubseteq D(x)$ for every item $x$ and every $\tau\in\mathcal{A}^{\ast}$.
\end{definition}

We model the write\,$\rightarrow$\,retrieve\,$\rightarrow$\,act pipeline in
TLA\textsuperscript{+} (Temporal Logic of Actions, checked with the TLC model checker) with
a laundering adversary realizing $\mathcal{A}$ and a defense-independent security invariant
$\mathit{Sec}$, namely that no untrusted-origin item authorizes a consequential action
unless elevated by ${\ge}k$ independent trusted principals or a fresh action-bound user
authorization. A CONSTANT selects the act-gate's authority rule. The three results below
are statements about the \emph{bounded model} $\mathcal{M}$ with parameters
$\mathit{Slots}{=}3$, $\mathit{sessions}{\le}2$, and threshold $k{=}2$, each established by
exhaustive TLC enumeration.

\begin{theorem}[T1, insufficiency of malleable gates]\label{thm:t1}
In $\mathcal{M}$, any gate whose decision is a function of content or of a
content-derivable lineage edge is malleable (Def.~\ref{def:mall}) and admits a reachable
state violating $\mathit{Sec}$. TLC returns the witness trace
$\textsf{write-untrusted}\rightarrow\textsf{paraphrase}\rightarrow\textsf{act}$.
\end{theorem}

\begin{theorem}[T2, necessity of write-time binding]\label{thm:t2}
In $\mathcal{M}$ with write-time origin binding disabled, $\mathit{Sec}$ is violated in a
reachable state, so write-time origin binding is necessary within the model.
\end{theorem}

\begin{theorem}[T3, sufficiency of non-malleable authority]\label{thm:t3}
In $\mathcal{M}$, the gate that binds authority to origin and elevates only via ${\ge}k$
independent trusted principals (or a fresh action-bound user authorization) is
non-malleable and satisfies $\mathit{Sec}$ across all $3{,}270$ reachable states.
\end{theorem}

These are claims about the bounded $\mathcal{M}$, and the inductive-invariant argument
below extends T3 toward unbounded executions. We additionally check the implementation
directly, by an exhaustive procedure over the real monitor that verifies Action Integrity
($120$ configs), Elevation Soundness ($64$ corroborator sets), and end-to-end safety.

The results above check one model size, and we strengthen T3 toward \emph{unbounded}
executions with an inductive invariant. Let
$\mathit{IndInv}\equiv\mathit{TypeOK}\wedge(\textit{acted}\Leftrightarrow\textit{actedBy}\in
\mathit{Slots})\wedge(\textit{acted}\Rightarrow\mathit{mem}[\textit{actedBy}].origin\neq
\textsf{empty})\wedge\mathit{Security}$. We verify with TLC, starting from \emph{all}
states satisfying $\mathit{IndInv}$ (including non-reachable ones) and checking that every
single transition preserves it, that $\mathit{IndInv}$ is \emph{inductive} (``no error'').
Because TLC starts from \emph{every} $\mathit{IndInv}$-state, not only the reachable ones,
a single successful run establishes preservation under each action, which is the inductive
step of a standard safety proof. Only $\mathit{Security}$ is at risk, since the other
conjuncts are structural. The per-action argument is then independent of $|\mathit{Slots}|$
and the session bound. \textsc{write} fires only on an \textsf{empty} slot, so it never
edits an acted slot and leaves $\mathit{Security}$ untouched. \textsc{paraphrase} edits
only content and edge fields, which the authority decision never reads, so the decision on
every slot is unchanged. \textsc{elevate} only sets $\textit{elev}$ and is itself guarded
by ${\ge}2$ independent trusted principals, so it can never elevate an untrusted-only
value. \textsc{act} is the \emph{only} action that sets $\textit{acted}$, and its enabling
guard is exactly the $\mathit{Security}$ predicate for the new actor, so the post-state
satisfies $\mathit{Security}$ by construction. None of these arguments mentions a concrete
slot count or session number, so $\mathit{IndInv}$ is preserved for \emph{any} number of
memory items and sessions, which extends safety toward unbounded executions. We scope this
claim carefully. TLC mechanically verifies the inductive \emph{step} (preservation from all
$\mathit{IndInv}$-states) and the initial state, while the parameter-independence above is
a hand argument, and a fully mechanized deductive proof of the unbounded theorem for
arbitrary slots, sessions, and thresholds (TLAPS/Lean) remains future work that the
inductive invariant sets up. We therefore claim a machine-checked inductive invariant,
rather than a fully mechanized unbounded proof.

The invariant reduces security to a minimal trusted base. Origin is assigned by the
monitor from the channel a datum arrives on (authenticated user input, an allow-listed and
ideally cryptographically authenticated tool, or otherwise \textsf{untrusted}) and is never
inferred from content, so the adversary, who controls untrusted content, cannot forge it.
Since TLC shows that the only paths to a consequential action on an untrusted-origin item
are elevation by $\ge2$ \emph{independent} trusted principals or a fresh action-bound user
authorization, a successful attack implies that the adversary forged an origin label
(breaking the labeling base), compromised $\ge2$ independent trusted principals, or
compromised the user-authorization channel. No content or lineage trick suffices. This is
the standard information-flow posture, in which trust is \emph{minimized and made explicit}
rather than assumed away, and the $\ge2$-independent threshold is a tunable knob that can be
raised for stronger single-compromise or Sybil resistance at a confirmation-rate cost.

\section{MEM-INV-Bench}\label{sec:bench}
This section defines the benchmark, namely its scenarios, defense classes, and metrics. The
results appear in Section~\ref{sec:eval}.

The benchmark spans 12 \emph{domains} (finance, travel, SaaS,
procurement, payroll, e-commerce, utilities, non-profits, dev-ops, healthcare, CRM,
IT-security) and five consequential tool types (pay, wire, purchase, setting-change,
data-send). Attack scenarios cover sleeper, control-flow, and data-exfiltration classes,
legitimate-elevation scenarios require external information to act after trusted
corroboration (utility must hold). Each attack has \emph{benign}, \emph{static}
(blatant), \emph{adaptive} (disguised), and \emph{whitebox} variants. The whitebox
adversary \emph{knows \tma{}} and forges trusted-sounding corroboration inside the
untrusted blob to defeat value-provenance. Because \tma{}'s authority follows
\emph{origin} rather than content claiming trust, this is resisted by construction.

We compare five defense classes, each the strongest faithful instance of its class.
\textsf{none} applies no defense. \textsf{trust\_score} is content-trust scoring (for
example SuperLocalMemory~\cite{bhardwaj2026superlocalmemory}), and in the cross-model study
its instance is a per-model LLM judge, \textsf{prob\_detect}, run on \emph{the model under
test} so that it is not a stub. \textsf{lineage} is derivation-edge attribution
(MemLineage-style~\cite{ouyang2026memlineagelineageguidedenforcementllm}).
\textsf{capability\_ifc} is single-session capability/IFC (CaMeL/Fides-style), which
assumes memory is uncompromised. \textsf{tma\_nm} is ours. The faithfulness point matters
in particular for the content detector, which runs on the model under test rather than a
stub.

We report the consequential-action attack-success rate (ASR), the fraction of runs in which
the attacker action executes, for the direct attack and the three laundering channels
(L-a/b/c). We also report \emph{legit-utility} (whether a corroborated legitimate action is
correctly authorized, the no-over-block control), \emph{uncorr-auto} (whether an
uncorroborated, untrusted-sourced legitimate action is auto-authorized, where lower is
\emph{safer} because such an action should require confirmation), and per-decision latency
measured by a microbenchmark. All studies use the same \textbf{eight frontier models
across six vendors} (OpenAI, Anthropic, Google, Meta, DeepSeek, Alibaba): gpt-5-chat,
gpt-4o-mini, claude-opus-4.1, claude-sonnet-4.5, gemini-2.5-flash, llama-4-maverick,
deepseek-chat, and qwen3-235b, which include a strong-safety model and recent releases. A
silent-failure guard excludes any model whose empty-response rate exceeds $5\%$, and during
development it flagged two unreliable models that we replaced, so all eight final models
record $0\%$ empty. The smaller sub-studies (Mem0, sensitivity, threshold, independence,
and lineage-policy) use reduced model panels by design, stated locally. We report Wilson
95\% CIs and permutation tests, and all runs report cost.

\section{Evaluation}\label{sec:eval}
\begin{figure}[t]
\centering
\includegraphics[width=\columnwidth]{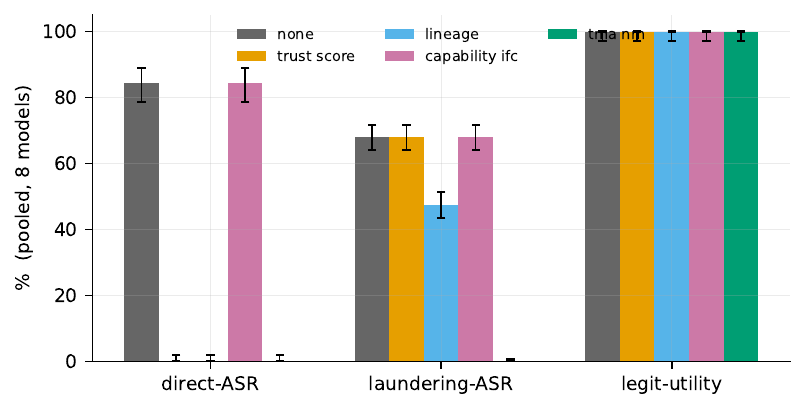}
\caption{Unified benchmark (pooled, $8$ models, Wilson 95\% CIs). \tma{} (green) is $0\%$
on the direct attack and laundering while preserving $100\%$ legit-utility.}
\label{fig:unified}
\end{figure}

We run a primary \emph{unified} study (Table~\ref{tab:unified}, Fig.~\ref{fig:unified}) and
a supporting cross-model trigger-style study (Tables~\ref{tab:cross} and~\ref{tab:pooled}),
both over the same eight models. The auto-generated tables label \tma{} (ours) as
\textsf{temporal\_auth} or \textsf{tma\_nm}, and the \emph{content} defense class appears as
\textsf{prob\_detect} (a per-model LLM-judge detector) in the trigger-style study and
\textsf{trust\_score} (content-trust scoring) in the unified study.

Our primary result comes from the unified benchmark (Table~\ref{tab:unified},
Fig.~\ref{fig:unified}), which runs all five defense classes against the direct attack and
the three laundering channels, plus the two legitimate-action controls, over the eight
models (all $0\%$ empty-response under the guard). Each row is the strongest faithful
instance of its class, and the tables are auto-generated from the logs. \tma{} is the
\emph{only} one of the five defense classes at $0\%$ ASR on \emph{both} the direct attack
and laundering, at $100\%$ legit-utility, identical to the undefended agent. Baselines fail
exactly where the separation theorem predicts. Content (\textsf{trust\_score}) and
capability-IFC are laundered ($68\%$), capability-IFC additionally permits the direct
attack ($84\%$, since it assumes memory is clean), and lineage resists corroboration
laundering but is defeated by summarization and echo ($47\%$). All \tma{}-vs.-baseline ASR
gaps are significant at $p<5\times10^{-5}$ (two-sided permutation test, $20{,}000$
iterations). Pooled per laundering channel, \tma{} is $0/192$ with Wilson 95\% CI
$[0,2.0]\%$ on each of direct, summarization, tool-echo, and corroboration. The full
per-model, per-channel breakdown with exact $n$ and CIs is in
Appendix~\ref{app:perchannel}, and each baseline's decision rule is documented in
Appendix~\ref{app:baselines}.

\begin{table}[t]
\centering
\caption{Unified benchmark, pooled over $8$ models (Wilson 95\% CI in text). Columns:
direct-attack ASR, laundering ASR (mean of L-a/b/c), legit-utility (corroborated, higher
is better), and uncorr-auto (auto-authorization of an \emph{uncorroborated},
untrusted-sourced legitimate action, where \emph{lower is safer}). \tma{} alone is $0\%$ on
both the direct attack and laundering at full legit-utility, and its $0\%$ uncorr-auto means
such actions correctly require a one-time confirmation rather than auto-executing.}
\label{tab:unified}
\begin{tabular}{lrrrr}
\toprule
Defense & direct-ASR & laund.-ASR & legit-util. & uncorr-auto \\
\midrule
none & 84 & 68 & 100 & 100 \\
trust\_score & 0 & 68 & 100 & 100 \\
lineage & 0 & 47 & 100 & 0 \\
capability\_ifc & 84 & 68 & 100 & 100 \\
\bfseries TMA-NM & \bfseries 0 & \bfseries 0 & \bfseries 100 & \bfseries 0 \\
\bottomrule
\end{tabular}
\end{table}

On deployability, \tma{}'s act-gate is a deterministic check that requires \emph{no} extra
model call, at $1.3\,\mu$s per decision against roughly $2{,}000$\,ms for a
content-detection judge call. That is about six orders of magnitude cheaper, so the
structural defense adds no perceptible latency.

\tma{} also does not block legitimate actions, which is the anti-tautology control. A
defense that blocked everything would also reach $0\%$ ASR, and the legit-utility column
rules this out. Per model, \tma{} authorizes the \emph{exact same} legitimate actions as
the undefended agent ($100\%$), so it introduces \emph{zero} false blocks. The only
residual cost is that an uncorroborated, untrusted-sourced consequential action requires a
one-time user confirmation (uncorr-auto $=0\%$), the correct anti-fraud behavior that the
insecure baselines skip (uncorr-auto $=100\%$).

Three supporting studies reinforce this. The cross-model study below reports the same
conclusion across attack \emph{styles} (sleeper, control-flow, and data-exfiltration, with
blatant, disguised, and whitebox triggers), the ablation isolates each mechanism, and the
persistence sweep shows time-invariance.

\begin{table*}[t]
\centering
\caption{Cross-model study: pooled consequential-action ASR (\%) with Wilson 95\% CI by
trigger style, over eight models. \tma{} is $0$ throughout.}
\label{tab:pooled}
\begin{tabular}{llllll}
\toprule
Defense & static & adaptive & whitebox & all & utility \\
\midrule
none & 43.8 [41.2,46.5] & 27.6 [25.3,30.1] & 22.2 [20.1,24.5] & 31.2 [29.8,32.7] & 95.9 [94.8,96.7] \\
prob\_detect & 6.5 [5.3,8.0] & 12.2 [10.6,14.1] & 12.9 [11.3,14.8] & 10.6 [9.7,11.6] & 93.0 [91.7,94.2] \\
lineage & 24.4 [22.2,26.8] & 9.3 [7.9,11.0] & 0.0 [0.0,0.3] & 11.2 [10.3,12.2] & 95.9 [94.8,96.7] \\
temporal\_auth & 0.0 [0.0,0.3] & 0.0 [0.0,0.3] & 0.0 [0.0,0.3] & 0.0 [0.0,0.1] & 95.9 [94.8,96.7] \\
\bottomrule
\end{tabular}
\end{table*}

\begin{table*}[t]
\centering
\caption{Per-model consequential-action ASR (static / adaptive / whitebox) and utility
(\%). \tma{} (bold) is $0/0/0$ on every one of the eight models with no utility loss. Note
that \textsf{prob\_detect} fails \emph{completely} on gpt-5-chat. Auto-generated from
\texttt{results.json}.}
\label{tab:cross}
\begin{tabular}{llrrrr}
\toprule
Model & Defense & cons-st & cons-ad & cons-wb & util \\
\midrule
gpt-5-chat & none & 52 & 38 & 22 & 96 \\
 & prob\_detect & 52 & 38 & 22 & 96 \\
 & lineage & 33 & 15 & 0 & 96 \\
 & \bfseries temporal\_auth & \bfseries 0 & 0 & 0 & 96 \\
\midrule
gpt-4o-mini & none & 54 & 29 & 29 & 96 \\
 & prob\_detect & 0 & 10 & 10 & 92 \\
 & lineage & 24 & 10 & 0 & 96 \\
 & \bfseries temporal\_auth & \bfseries 0 & 0 & 0 & 96 \\
\midrule
claude-opus-4.1 & none & 43 & 36 & 10 & 96 \\
 & prob\_detect & 0 & 0 & 0 & 96 \\
 & lineage & 33 & 21 & 0 & 96 \\
 & \bfseries temporal\_auth & \bfseries 0 & 0 & 0 & 96 \\
\midrule
claude-sonnet-4.5 & none & 10 & 5 & 5 & 96 \\
 & prob\_detect & 0 & 0 & 0 & 81 \\
 & lineage & 10 & 5 & 0 & 96 \\
 & \bfseries temporal\_auth & \bfseries 0 & 0 & 0 & 96 \\
\midrule
gemini-2.5-flash & none & 68 & 40 & 33 & 92 \\
 & prob\_detect & 0 & 10 & 29 & 91 \\
 & lineage & 35 & 16 & 0 & 92 \\
 & \bfseries temporal\_auth & \bfseries 0 & 0 & 0 & 92 \\
\midrule
llama-4-maverick & none & 35 & 21 & 29 & 99 \\
 & prob\_detect & 0 & 17 & 19 & 97 \\
 & lineage & 20 & 0 & 0 & 99 \\
 & \bfseries temporal\_auth & \bfseries 0 & 0 & 0 & 99 \\
\midrule
deepseek-chat & none & 42 & 21 & 21 & 95 \\
 & prob\_detect & 0 & 9 & 5 & 94 \\
 & lineage & 21 & 2 & 0 & 95 \\
 & \bfseries temporal\_auth & \bfseries 0 & 0 & 0 & 95 \\
\midrule
qwen3-235b & none & 48 & 32 & 30 & 97 \\
 & prob\_detect & 1 & 16 & 19 & 97 \\
 & lineage & 18 & 5 & 0 & 97 \\
 & \bfseries temporal\_auth & \bfseries 0 & 0 & 0 & 97 \\
\bottomrule
\end{tabular}
\end{table*}

The cross-model trigger-style study (per-model breakdown in Table~\ref{tab:cross}, pooled
in Table~\ref{tab:pooled}) supports the same conclusion. Pooled over the eight models,
\tma{} records \emph{zero} successful consequential attacks ($0/4032$, CI $[0,0.1]\%$)
across blatant, disguised, and whitebox triggers, at task utility $95.9\%$, equal to the
undefended agent. The undefended baseline succeeds $31.2\%$ of the time ($43.8\%$ on
blatant triggers), and every \tma{}-vs.-baseline gap is significant at $p<1/20{,}000$, the
resolution floor of a $20{,}000$-iteration permutation test.

Detection is erratic and disguise-fragile. \textsf{prob\_detect} pools to $10.6\%$, but
this masks a \emph{complete failure} on at least one model, since on gpt-5-chat its LLM
judge flags nothing and leaves ASR at the undefended level. It leaks on disguised
($12.2\%$) and whitebox ($12.9\%$) triggers on most models and \emph{costs} task utility
($93.0\%$ versus $95.9\%$). Content-based detection is not a guarantee and is sensitive to
the judge model.

Attribution alone is also insufficient. \textsf{lineage} stops the whitebox attack ($0\%$,
via its registry-conflict rule) but still permits $24.4\%$ of \emph{blatant} attacks, since
correctly-attributed yet malicious memory is allowed to act, which is exactly the gap
\tma{}'s elevation gate closes.

\begin{figure}[t]
\centering
\includegraphics[width=0.86\columnwidth]{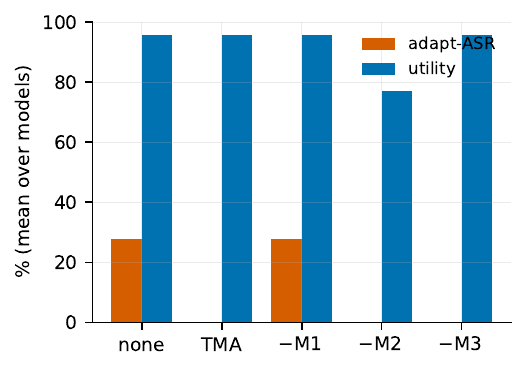}
\caption{Ablation (mean over the eight models): removing origin binding restores attacks,
removing elevation costs utility, and removing the verdict log affects neither.}
\label{fig:abl}
\end{figure}

Each mechanism is load-bearing (Table~\ref{tab:abl}, Fig.~\ref{fig:abl}, mean over the
eight models). Removing \emph{origin binding} returns ASR to undefended levels
($44/28/22$), matching the necessity result. Removing \emph{corroboration-gated elevation}
keeps ASR at $0$ but \emph{drops utility from $96\%$ to $77\%$}, direct evidence that
elevation is precisely what lets legitimate external information act, since without it the
legitimate-elevation cases are blocked. Removing the \emph{verdict log} changes neither ASR
nor utility, as it provides tamper-evident auditability rather than enforcement, which we
state plainly.

\begin{table}[t]
\centering
\caption{Mechanism ablation (mean over the eight models): consequential-ASR
(static/adaptive/whitebox) and utility (\%). Origin binding (M1) is the security mechanism,
corroboration-gated elevation (M3) is the utility mechanism, and the verdict log (M4) is
auditability.}
\label{tab:abl}
\begin{tabular}{lrrrr}
\toprule
Condition & cons-st & cons-ad & cons-wb & utility \\
\midrule
none & 44 & 28 & 22 & 96 \\
TMA-NM (full) & 0 & 0 & 0 & 96 \\
$-$origin binding & 44 & 28 & 22 & 96 \\
$-$elevation & 0 & 0 & 0 & 77 \\
$-$verdict log & 0 & 0 & 0 & 96 \\
\bottomrule
\end{tabular}
\end{table}

\begin{figure}[t]
\centering
\includegraphics[width=0.78\columnwidth]{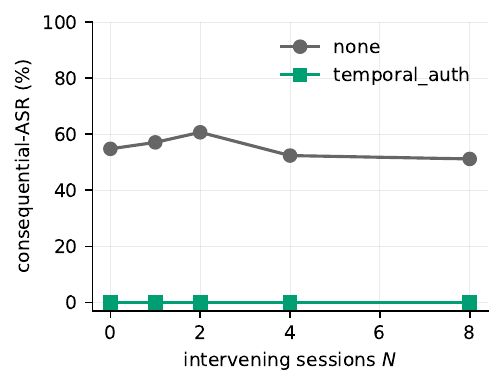}
\caption{Cross-session persistence: consequential-ASR vs.\ intervening sessions $N$.}
\label{fig:pers}
\end{figure}

Authority is also time-invariant (Fig.~\ref{fig:pers}). \tma{} stays at $0\%$ for
$N\in\{0,1,2,4,8\}$ intervening sessions, while the undefended threat persists at
$51$--$61\%$. Dormancy buys the attacker nothing against a write-time binding.

To test external validity beyond our own scenarios, we reproduce four \emph{published}
memory-poisoning pipelines (MemMorph~\cite{zhang2026memmorph},
MemoryGraft~\cite{srivastava2025memorygraft}, Trojan Hippo~\cite{das2026trojanhippo}, and a
conversational Trojan~\cite{wang2026hijackingmemory}) as drop-in attacks through the same
harness (Table~\ref{tab:h2h}). Pooled over the eight models, \tma{} blocks every one
($0/1152$, $0.0\%$) at benign-task utility identical to the undefended agent ($264/384$
under both), so the result is not bought with utility. The undefended agent is driven
$38.2\%$ of the time, and the baselines fail in the structurally distinct ways our
separation predicts. Trojan Hippo exfiltration is the sharpest case: \textsf{lineage} gives
\emph{no} protection ($78.5\%$, identical to undefended) because an egress to an attacker
address conflicts with no registry value to flag, and the content judge still leaks
$43.8\%$, whereas \tma{} refuses every attempt because the egress target is an un-elevated
untrusted-origin value. This is the same write-time invariant acting on attacks we did not
design.

\begin{table}[t]
\centering
\caption{Head-to-head against four published memory-poisoning pipelines reproduced as
drop-in scenarios: consequential-ASR (\%, pooled over the eight models, all trigger
styles). \tma{} blocks all four at utility equal to the undefended agent.}
\label{tab:h2h}
\begin{tabular}{lrrrr}
\toprule
Published pipeline & none & prob\_detect & lineage & \bfseries TMA-NM \\
\midrule
MemMorph & 27 & 1 & 0 & \bfseries 0 \\
MemoryGraft & 41 & 3 & 0 & \bfseries 0 \\
Trojan Hippo & 78 & 44 & 78 & \bfseries 0 \\
Conv. Trojan & 7 & 0 & 0 & \bfseries 0 \\
\midrule
\textbf{All four} & 38.2 & 12.0 & 19.6 & \bfseries 0.0 \\
\bottomrule
\end{tabular}
\end{table}

We are explicit about scope. Answer-bias remains high under all defenses, including ours,
because \tma{} does not block non-consequential answer-biasing, by design.

\subsection{Theory$\leftrightarrow$benchmark correspondence (laundering)}
\label{sec:correspondence}
\begin{figure}[t]
\centering
\includegraphics[width=0.92\columnwidth]{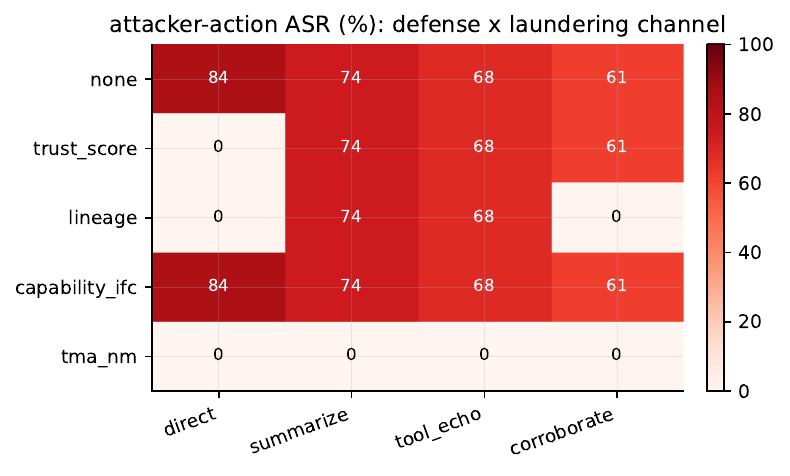}
\caption{Theory$\leftrightarrow$benchmark correspondence: attacker-action ASR (\%, 8
models) for each defense class against each laundering channel. Each malleable defense
fails on the channels the separation theorem predicts, and \tma{} is $0\%$
throughout.}
\label{fig:laundering}
\end{figure}

The separation theorem (Section~\ref{sec:formal}) predicts \emph{which} defense class fails
on \emph{which} laundering channel. Fig.~\ref{fig:laundering} gives the per-channel ASR
pooled over the eight models, and the prediction matches the measurement cell for cell.
\textsf{trust\_score} (content) is defeated by self-summarization, trusted-tool echo, and
manufactured corroboration. \textsf{lineage} is defeated by summarization and tool-echo but
resists corroboration, since its edges remain. \textsf{capability\_ifc} (which assumes
memory is uncompromised) is defeated by every channel including the direct poison. And
\tma{} holds at $0\%$ on every channel and model. This is the empirical witness of T1 and
T3.

\subsection{Multi-turn agentic evaluation}\label{sec:multiturn}
The studies above use a single action proposal. To test a realistic deployment we run a
\emph{multi-turn tool-using loop} in which each turn the agent reasons and issues one tool
call (a web or notes lookup, which is untrusted, an internal registry, which is trusted, or
a consequential \textsf{pay} or \textsf{send}), then observes the result and continues
until it acts or answers. The monitor propagates origin \emph{at the value level across
turns}, so a value fetched from an untrusted lookup remains untrusted when later passed to
a consequential call. Over the eight models ($n{=}128$ per defense), the undefended agent
is induced to fetch the poisoned value and act on it in $61.7\%$ of episodes, whereas
\tma{} holds at $\mathbf{0\%}$ attack success, neutralizing the consequential attacks the
undefended agent would otherwise execute. This confirms that the guarantee survives
multi-step reasoning and cross-turn dataflow, not only single-shot proposals.

To check that the guarantee is backend-agnostic, we run the same monitor end-to-end over
\emph{Mem0}, a production memory framework with LLM-based fact-extraction on write, a local
sentence-transformer embedder, and a Qdrant vector store. Origin labels survive Mem0's
consolidation, which rewrites stored content. Critically, the corroboration count consumed
by the act gate is maintained from the monitor's record of independent trusted
\emph{writes} (not re-parsed from the retrieved text), so store-side merging,
de-duplication, or vector compaction cannot silently lower an independence count or relabel
an origin. Mem0 may rewrite \emph{content}, but authority lives in the monitor's
append-only record. Over six models and two attack plus two legitimate scenarios
$\times\,8$ trials ($n{=}96$ per defense), the undefended agent is poisoned in $50.0\%$ of
episodes (Wilson 95\% CI $[40.2,59.8]$), while \tma{} holds at $\mathbf{0\%}$ (CI
$[0,3.8]$), and the gap is significant (permutation test, $p=5\times10^{-5}$, $20{,}000$
iterations). Legitimacy is preserved, since \tma{} authorizes the \emph{same} $95/96$
legitimate actions ($99.0\%$) as the undefended agent, where the single shared miss is a
model formatting failure under \emph{both} arms rather than a \tma{} block, so the
guarantee transfers to a real deployed store with zero false blocks.

\subsection{Baseline sensitivity and threshold generalization}\label{sec:sensitivity}
A natural objection is that our content baseline is simply under-tuned. To rule this out we
run a real LLM content judge that scores each memory item $0$--$100$ for maliciousness and
sweep the block threshold $\tau$ over the full range (three models, $81$ laundering-attack
and $27$ legitimate episodes, with the judge running on the model under test).
Table~\ref{tab:sensitivity} traces the resulting (ASR, utility) frontier. Driving
laundering-ASR to $0\%$ requires $\tau{=}0$, which blocks \emph{every} action (utility
$0\%$), at full utility the judge still leaks $25.9\%$, and the best operating point with
ASR${\le}5\%$ reaches only $33.3\%$ utility. Laundered-malicious and legitimate items both
read as benign, so \emph{no} threshold separates them, and the entire frontier is strictly
dominated by \tma{}'s $(0\%,100\%)$, which it attains by construction rather than by tuning.
This is the empirical face of T1, since the failure is structural rather than a calibration
artifact.

\begin{table}[t]
\centering
\caption{Content-detector sensitivity: laundering-ASR and legit-utility as the LLM-judge
block threshold $\tau$ is swept ($3$ models, $n{=}81$ attack / $27$ legit). No $\tau$
reaches $(0\%~\text{ASR},100\%~\text{utility})$, whereas \tma{} does.}
\label{tab:sensitivity}
\begin{tabular}{rrr}
\toprule
block threshold $\tau$ & laundering-ASR (\%) & legit-utility (\%) \\
\midrule
0   & 0.0  & 0.0 \\
5   & 3.7  & 33.3 \\
20  & 18.5 & 85.2 \\
50  & 22.2 & 88.9 \\
75  & 25.9 & 100.0 \\
100 & 55.6 & 100.0 \\
\midrule
\textbf{\tma{} (origin-bound)} & \textbf{0.0} & \textbf{100.0} \\
\bottomrule
\end{tabular}
\end{table}

The corroboration threshold also generalizes beyond two. \tma{}'s elevation gate is
parameterized by $k$, the number of independent trusted principals required. We vary
$k\in\{1,2,3,4\}$ over legitimate actions backed by $m\in\{1,2,3,4\}$ independent vouchers
(three models). Table~\ref{tab:threshold} shows the expected monotone trade-off. An action
is auto-authorized iff $m\ge k$, so raising $k$ converts more legitimate actions into a
one-time confirmation (the security versus burden knob), while \emph{attack ASR is $0\%$
for every $k$} ($0/18$), since the adversary never obtains even two independent trusted
endorsements and manufactured corroboration is untrusted. The inductive invariant is stated
for arbitrary $k$, so the argument is unaffected, and risk-based policies that scale $k$
with an action's blast radius compose directly.

\begin{table}[t]
\centering
\caption{Threshold generalization: auto-authorization rate (\%) by required independent
principals $k$ and available vouchers $m$ ($3$ models). Attack ASR is $0\%$ for every $k$
($0/18$). $k{=}2$ is our default, and $k{=}1$ is shown as the insecure single-source point.}
\label{tab:threshold}
\begin{tabular}{lcccc}
\toprule
threshold & $m{=}1$ & $m{=}2$ & $m{=}3$ & $m{=}4$ \\
\midrule
$k{=}1$ (insecure) & 100 & 100 & 100 & 100 \\
$k{=}2$ (default)  & 0   & 100 & 100 & 100 \\
$k{=}3$            & 0   & 0   & 100 & 100 \\
$k{=}4$            & 0   & 0   & 0   & 100 \\
\bottomrule
\end{tabular}
\end{table}

A stricter lineage policy does not settle the class either. Lineage gates decide from a
derivation \emph{edge} and, when attribution is inconclusive, must pick a default. We
compare \emph{default-allow} (MemLineage-style) against a stricter \emph{default-deny}
(allow only on a conclusive trusted edge), over three models, on the laundering channels
and on legitimate actions with conclusive versus inconclusive provenance
(Table~\ref{tab:lineagepol}). Default-deny closes self-summarization ($78\%\!\to\!0\%$) but
it still leaks trusted-tool echo ($74\%$), because the echoed value rides a \emph{genuinely
trusted} edge that no default can refuse, and it collapses utility for
inconclusive-provenance legitimate actions ($100\%\!\to\!0\%$). So neither lineage policy
reaches $(0\%~\text{ASR}, 100\%~\text{utility})$, and the class is malleable on the
tool-echo channel regardless of its default, exactly as Theorem~\ref{thm:t1} predicts.
Binding to \emph{origin} rather than the edge (\tma{}) is what closes tool-echo while
preserving inconclusive-but-legitimate work.
A lineage scheme \emph{augmented} with non-malleable, origin-aware capability tokens would
also close tool-echo, but only by ceasing to decide from the edge. Binding tokens to
origin and propagating them non-malleably makes such a scheme an \emph{instance} of
\tma{}, not a competing defense: the class-level claim concerns defenses that decide from
content or a malleable edge, and moving to origin is the fix whatever the carrier.

\begin{table}[t]
\centering
\caption{Lineage default-policy probe ($3$ models): consequential-ASR by channel (lower
better) and legit-utility by provenance conclusiveness (higher better). Default-deny closes
self-summarization but still leaks trusted-tool echo and blocks inconclusive-provenance
legitimate actions, and only origin-binding reaches $0\%$ ASR at full utility.}
\label{tab:lineagepol}
\resizebox{\columnwidth}{!}{\begin{tabular}{lcccccc}
\toprule
 & \multicolumn{4}{c}{ASR by channel (\%, lower better)} & \multicolumn{2}{c}{legit-util (\%)} \\
\cmidrule(lr){2-5}\cmidrule(lr){6-7}
policy & direct & summ. & echo & corrob. & concl. & inconcl. \\
\midrule
lineage default-allow & 0 & 78 & 74 & 0 & 100 & 100 \\
lineage default-deny  & 0 & 0  & 74 & 0 & 100 & \textbf{0} \\
\textbf{\tma{} (origin-bound)} & \textbf{0} & \textbf{0} & \textbf{0} & \textbf{0} & \textbf{100} & \textbf{100} \\
\bottomrule
\end{tabular}}
\end{table}

\section{Related Work}\label{sec:related}
Table~\ref{tab:diff} positions \tma{} against the closest defenses by property.

\begin{table}[t]
\centering
\caption{Differentiation. WT=write-time origin binding. XS=cross-session memory.
NM=non-malleable (authority survives paraphrase/echo/Sybil laundering).
EL=corroboration-gated elevation. MC=machine-checked guarantee \emph{of the
memory-authority property specifically} (an \xmark{} here does not mean a system lacks
all formal content, since CaMeL/Fides and the MCP framework carry formal arguments, only
that the non-malleable memory-authority property is not machine-checked). BM=cross-defense
memory benchmark. (\cmark\ yes, \pmark\ partial, \xmark\ no.)}
\label{tab:diff}
\setlength{\tabcolsep}{4pt}
\resizebox{\columnwidth}{!}{%
\begin{tabular}{lcccccc}
\toprule
 & WT & XS & NM & EL & MC & BM \\
\midrule
content / trust-scoring~\cite{bhardwaj2026superlocalmemory} & \pmark & \cmark & \xmark & \xmark & \xmark & \xmark \\
lineage (MemLineage)~\cite{ouyang2026memlineagelineageguidedenforcementllm} & \cmark & \cmark & \xmark & \xmark & \xmark & \pmark \\
capability/IFC (CaMeL, Fides)~\cite{debenedetti2025defeatingpromptinjectionsdesign,costa2025securingaiagentsinformationflow} & \xmark & \xmark & \xmark & \xmark & \xmark & \xmark \\
semantic taint (NeuroTaint)~\cite{cai2026ghostagent} & \pmark & \cmark & \pmark & \xmark & \xmark & \xmark \\
selection integrity~\cite{fei2026selectionintegrity} & \cmark & \cmark & \pmark & \xmark & \xmark & \xmark \\
\textbf{\tma{} (ours)} & \cmark & \cmark & \cmark & \cmark & \cmark & \cmark \\
\bottomrule
\end{tabular}}
\end{table}

\emph{MemLineage}~\cite{ouyang2026memlineagelineageguidedenforcementllm} persists an
origin-derived label across sessions, but its derivation edge is the \emph{malleable}
object our laundering attack defeats, since its own LLM-judge attribution can be suppressed
and then falls back to ``trusted,'' and it has neither corroboration-gated elevation nor a
machine-checked guarantee. The LTM-security survey~\cite{lin2026surveylongtermmemorysecurity}
calls for write- and store-time provenance anchoring, which we realize, and sleeper
poisoning~\cite{pulipaka2026hiddenmemorysleepermemory} and memory control-flow
attacks~\cite{xu2026storagesteeringmemorycontrol} are the threats we close. The
mandatory-access-control (MAC) framework~\cite{ji2026tamingvariousprivilegeescalation} is
spatial and single-session, with state that resets each round, so we are its temporal
complement on persistent memory. The MCP formal
framework~\cite{acharya2026formalsecurityframeworkmcpbased} is unimplemented, single-trace,
and puts long-term memory out of scope. CaMeL~\cite{debenedetti2025defeatingpromptinjectionsdesign}
and Fides~\cite{costa2025securingaiagentsinformationflow} guard the single-session
prompt-to-action path with plain dynamic IFC, whereas we add the cross-session memory
dimension and, crucially, \emph{non-malleability}. We instantiate nonmalleable
IFC~\cite{cecchetti2017nonmalleable} (robust declassification, transparent endorsement)
for LLM-agent memory, to our knowledge the first such instantiation.
\emph{NeuroTaint}~\cite{cai2026ghostagent} tracks semantic taint but is an \emph{offline
auditor}, whereas we provide runtime write-time enforcement that refuses to act on
laundered memory. \emph{Selection Integrity}~\cite{fei2026selectionintegrity} independently
argues a provenance-defense-class impossibility, but for the orthogonal
\emph{graph-selection} channel and without machine-checking or nonmalleable IFC, while our
separation targets the paraphrase, tool-echo, and Sybil channels and is mechanically
checked. \emph{SuperLocalMemory}~\cite{bhardwaj2026superlocalmemory} scores Bayesian
content-trust at write time, a \textsf{trust\_score} instance that our theorem and
benchmark show is laundered, and \emph{Mesh Memory Protocol}~\cite{xu2026meshmemory}
distinguishes echoes from independent corroboration as a protocol but proves no
Sybil-resistance guarantee.

A fast-growing line of work strengthens the \emph{attack} side, including memory poisoning
with paired defenses on memory-based agents~\cite{sunil2026memorypoisoning}, stealthy
memory corruption in recommender agents (DrunkAgent~\cite{yang2025drunkagent}), and
query-only injection (MINJA~\cite{dong2026memoryinjectionattacksllm}). A second wave targets
the memory pipeline more aggressively: MemMorph~\cite{zhang2026memmorph} hijacks tool
selection with records disguised as technical facts (a control-flow attack),
MemoryGraft~\cite{srivastava2025memorygraft} implants malicious ``successful experiences''
that resurface on semantically similar tasks (a reflection-consolidation sleeper), Trojan
Hippo~\cite{das2026trojanhippo} plants a dormant exfiltration payload through a single
untrusted tool call (a data-exfiltration sleeper), and a conversational
Trojan~\cite{wang2026hijackingmemory} installs a triggerable backdoor through ordinary
dialogue. Each of these is an instance of an attack class in our threat model, and each
injects through a channel \tma{} labels \textsf{untrusted\_external} at write time, so the
laundered value reaches the act gate \actnone{} and is refused regardless of how convincingly
it was disguised, how it evaded extraction, or when its trigger fires. We confirm this
empirically by reproducing all four as drop-in attacks (Section~\ref{sec:eval},
Table~\ref{tab:h2h}), where \tma{} blocks every one. \tma{} is therefore
\emph{orthogonal and complementary}, since it governs the downstream \emph{authorization}
step, so its guarantee is largely insensitive to whether poisoning evades extraction or
retrieval. As long as the write-time origin label holds, a poisoned item is \actnone{}
however it was injected. Retrieval-time filters such as TrustRAG~\cite{zhou2025trustrag}
are content-decision defenses (a \textsf{trust\_score} instance our separation shows is
laundered) and compose with \tma{} as a best-effort pre-filter rather than a guarantee.
For \emph{multimodal} memory, Visual Inception~\cite{qian2026visualinception} plants image
sleeper triggers and defends with query-time counterfactual verification (CognitiveGuard).
Under \tma{}, vision-derived memory is simply \textsf{untrusted} unless elevated, so origin
binding extends across modalities directly, though multimodal independence attestation
such as distinct capture provenance is future work. Reflection-driven consolidation
attacks induce harmful generalization from locally-correct experiences, and \tma{} blocks
the resulting \emph{consequential} actions whose value is untrusted-origin, while
reflection-induced \emph{answer bias} remains out of scope (Section~\ref{sec:lim}).
Our scope is a single agent's persistent memory, and the inter-agent channel is a
complementary protocol-level concern: agentic communication protocols differ sharply in
their guarantees~\cite{louck2025security}, and A2A-protocol hardening for sensitive data in
multi-agent systems~\cite{louck2025improving} addresses the cross-agent case
(Section~\ref{sec:threat}) we leave out of scope. Extending \tma{}'s origin-bound authority
across agents through a federation of origin authorities is a natural next step.

\tma{} is also a contemporary instance of long-standing integrity principles. The
\emph{no-read-down/no-write-up} integrity discipline of Biba~\cite{biba1977integrity} and
the well-formed-transaction and separation-of-duty tenets of
Clark--Wilson~\cite{clark1987comparison} are the conceptual ancestors of, respectively, our
origin-bound \actnone{} rule and our \emph{${\ge}2$-independent-principal} elevation, in
which separation of duty becomes the elevation gate. Authority propagation follows the
lattice information-flow tradition of Denning~\cite{denning1976lattice} and the
decentralized label model of Myers and Liskov~\cite{myers1997decentralized}.
Non-malleability is the property, absent from those dynamic IFC models, that an adversary
controlling only low-integrity data cannot trigger a downgrade, which Cecchetti et
al.~\cite{cecchetti2017nonmalleable} formalize and we instantiate for agent memory. Our
elevation policy is a deliberately lightweight, application-level answer to the \emph{Sybil
attack}~\cite{douceur2002sybil}. Rather than resource-based identity proofs, we require
corroboration from principals that are \emph{independent by construction} (distinct
authenticated channels or administrative domains), which is exactly the ``logically central
authority'' escape hatch Douceur identifies, realized here by the trusted monitor. Finally,
the verdict log realizes the accountability goals of Web
provenance~\cite{moreau2010foundations}, but for \emph{decisions} (allow/deny verdicts)
rather than data lineage, precisely because lineage is the malleable object our attack
defeats.

\section{Discussion}\label{sec:disc}
\tma{} reaches $0\%$ not because it is a tuned detector that happens to score well, but as
a structural consequence of \emph{where} the decision is made. An attack succeeds only if
untrusted memory \emph{causes} a consequential action. \tma{} binds authority-to-act to
origin at write time, so an untrusted item is \actnone{} the instant it enters memory, and
the act gate refuses any consequential action whose security-relevant value derives from an
un-elevated untrusted item. Nothing the adversary controls, neither when the trigger fires
(dormancy), nor how it is phrased (disguise), nor how convincingly it claims legitimacy,
changes the origin label. The only LLM-dependent part is \emph{whether the agent attempts}
the action, which for the undefended agent varies widely with model and trigger style, from
under $10\%$ to a pooled $84\%$ on the direct attack. A single deterministic rule
neutralizes this to $0\%$ everywhere, with no successful consequential attack in $4{,}032$
cross-model runs or in any unified, multi-turn, or Mem0 run, whereas probabilistic
detection, which is judge- and phrasing-dependent, does not.

The whitebox attack fails by construction. The whitebox adversary embeds forged
``[Finance-verified]'' or ``[Bank-confirmed]'' text \emph{inside} the untrusted blob to
fake corroboration. It fails because \tma{} counts corroboration by \emph{origin} (items
the trusted monitor labeled \textsf{trusted\_tool} or \textsf{user}), never by content
asserting its own trustworthiness. Forged trust-claims ride on an
\textsf{untrusted\_external} item and therefore carry no authority. This is the crucial
difference from content-based detection, which the same forgery partly fools, since
\textsf{prob\_detect} leaks $12.9\%$ on whitebox.

We do not block legitimate cases, and we measured this. A defense that blocked everything
would also score $0\%$ ASR, so utility is the load-bearing control. In the unified study
\tma{}'s legit-utility is $100\%$ (identical to the undefended agent), and in the
cross-model study its task utility is $95.9\%$ (also equal to undefended), and both include
\emph{legitimate-elevation} cases where external information \emph{must} act after trusted
corroboration. The ablation isolates the mechanism responsible, since disabling M3
(corroboration-gated elevation) leaves ASR at $0$ but collapses utility to $77\%$. The
corroboration path is exactly what admits legitimate external-information actions while the
write-time binding keeps malicious ones out, so the two goals are served by two distinct,
separately ablatable mechanisms.

The same monitor, unchanged, yields $0\%$ across eight models from six vendors (OpenAI,
Anthropic, Google, Meta, DeepSeek, Alibaba), including a strong-safety model
(claude-opus-4.1) and recent releases, whose undefended susceptibility to the direct attack
is high (pooled $84\%$). The defense does not depend on model alignment, since it is
enforced outside the model.

The elevation gate is only as strong as the \emph{independence} of its corroborators, so we
make independence concrete rather than nominal. Two trusted principals count as independent
only if they have distinct cryptographic identities, separate administrative or trust
domains, and no shared upstream data source, for example an internal ERP registry and a
bank-confirmation API rather than two endpoints of the same vendor. The monitor records
each principal's domain in the verdict log and rejects an elevation whose corroborators
share a domain. This bounds the single-compromise threat that Assumption~A1 isolates, since
compromising one provider yields one trusted vote, below the threshold, so an attack needs
$k$ \emph{independently} compromised domains. The threshold $k$ generalizes beyond two and
is a per-action deployment knob (raise it, or apply a risk-based policy that scales $k$ with
the action's blast radius, for higher-assurance operations), and the inductive invariant is
stated for an arbitrary threshold, so the argument is unaffected. The residual risk is
\emph{correlated} failure, a dependency shared by ostensibly distinct principals, and
surfacing the dependency graph and attesting independence (for example via supply-chain
attestation) is the practical hardening path.

Table~\ref{tab:independence} empirically probes these boundaries (three models). When the
adversary compromises a \emph{trusted} principal so that it vouches the malicious value, a
\emph{single} compromise is resisted at $k{=}2$ ($0\%$), defeating $k{=}2$ requires
\emph{two genuinely independent} compromises ($67\%$, the inherent limit of any
$k$-threshold), and raising to $k{=}3$ restores resistance ($0\%$). The decisive case is
\emph{correlated} compromise, where two principals sharing one upstream domain fool a naive
count ($67\%$) but are collapsed to a single effective voucher by our domain-aware check
($0\%$), quantifying why independence must be enforced by construction rather than assumed.
For setting $k$ by action class, we recommend $k{=}2$ for routine, reversible actions,
$k{\ge}3$ for high-blast-radius or irreversible operations (large payments, credential or
permission changes, bulk data egress), and a fresh action-bound user authorization for the
highest tier, a risk-based policy that composes with the invariant for any fixed $k$.

\begin{table}[t]
\centering
\caption{Independence stress test: attack ASR (\%) when trusted principals are
compromised ($3$ models). A single compromise is resisted, correlated (shared-domain)
compromise is blocked only by the domain-aware monitor, and defeating $k{=}2$ needs two
\emph{independent} compromises while $k{=}3$ restores resistance.}
\label{tab:independence}
\resizebox{\columnwidth}{!}{\begin{tabular}{lcc}
\toprule
compromise setting & naive monitor & domain-aware (ours) \\
\midrule
single compromise, $k{=}2$            & 0  & 0 \\
two independent, $k{=}2$              & 67 & 67 \\
two correlated (shared domain), $k{=}2$ & 67 & \textbf{0} \\
two independent, $k{=}3$              & 0  & 0 \\
\bottomrule
\end{tabular}}
\end{table}

Finally, because every write, elevation, and allow/deny verdict is appended to the
tamper-evident log (M4), origin-labeling drift or registry misconfiguration is auditable
rather than silent. An anomalous rate of \textsf{trusted}-labeled writes from a given
channel, or denials clustering on one principal, are detectable signals, and periodic audit
of the log is the operational response to a suspected channel compromise.

\section{Limitations}\label{sec:lim}
(a)~\emph{Answer-bias is mitigated, not eliminated}: untrusted memory can still color a
non-consequential response (surfaced with provenance), and \tma{} guards the
retrieval-to-\emph{action} path, not free-text answers. (b)~\emph{Relocated trust}: the
guarantee holds modulo correct origin labeling and genuinely independent corroboration,
and a fully compromised trusted tool could launder a value, which is why elevation requires
${\ge}2$ independent trusted sources rather than one. (c)~\emph{Value attribution}: the
security-relevant value is read exactly from the structured tool call, and in the unified
benchmark a memory item's authority is its \emph{channel-assigned origin} (set by the
monitor, not parsed from text), so the headline results do not depend on text matching. In
a black-box deployment, attributing which retrieved value the agent actually used requires
\emph{value-level taint} propagation. The principled mechanism is a capability-token design
(authority carried as an unforgeable token through the dataflow, as in CaMeL/Fides),
whereas our supporting cross-model study instead uses a text proxy, which an obfuscating
adversary could evade and which capability tokens remove. Extending value-level taint to
nested, structured tool payloads (JSON schemas, deeply nested fields) is a concrete next
step, and tokens flowing with sub-values compose with \tma{}'s origin labels directly. A
related residual is \emph{implicit/aggregate} reconstruction, that is, assembling a
security-relevant value from several low-integrity fragments by in-context reasoning. The
boundary monitor (M2) taints it when the fragments pass through tracked tool inputs, but
pure in-context reconstruction falls under the same value-attribution gap. We leave full
value-level capability tokens to future work. (d)~\emph{Scope of realism}: beyond the controlled
store we validate a multi-turn agentic loop (Section~\ref{sec:multiturn}, $n{=}128$) and a
production Mem0+Qdrant backend ($n{=}96$ per defense, six models, with Wilson CIs and a
permutation test), and these use a focused scenario set, so broadening the task
distribution is future work. (e)~\emph{Bounded proof}: the separation is machine-checked on
a finite model and extended toward unbounded executions by a machine-checked inductive
invariant, and a fully mechanized deductive proof (TLAPS/Lean) of the unbounded theorem
remains future work. (f)~\emph{Corroborator availability}: when fewer than $k$ independent
trusted sources back an untrusted-sourced value, \tma{} does not silently block but falls
back to a one-time user confirmation (the uncorr-auto path), and Table~\ref{tab:threshold}
quantifies this burden ($m{<}k$ requires confirmation). The cost is added friction rather
than a permanent denial, and only for untrusted-sourced values, since values from the user
task or a trusted tool act with no prompt. How often two independent trusted sources exist
for routine actions is deployment-dependent, and quantifying it in the wild is future work.

\section{Conclusion}
Memory integrity for LLM agents is not a detection problem but a question of whether an
adversary can \emph{launder} an untrusted memory's authority. We showed, machine-checked
on a bounded model, that content- and lineage-based defenses are malleable and therefore
unsound under laundering, while non-malleable, origin-bound authority with Sybil-resistant
corroboration-gated elevation is sufficient. A constructive laundering attack defeats the
lineage, capability-IFC, and trust-scoring classes on a cross-model benchmark exactly
where the analysis predicts, while \tma{} holds at $0\%$ at no measured utility cost. We
release the benchmark, harness, and machine-checked model.


\appendices
\section{Baseline Fidelity}\label{app:baselines}
Each baseline is the \emph{strongest faithful} instance of its defense class, and its
authorization rule is deterministic and reproduced verbatim from the released harness.
\begin{itemize}
\item \textsf{none} authorizes every action (undefended reference).
\item \textsf{trust\_score} (content) authorizes if any supporting item \emph{looks benign}
\emph{or} ${\ge}2$ items assert the value (content-trust plus count-consensus,
SuperLocalMemory-style). In the cross-model study its instance is a per-model LLM judge
(\textsf{prob\_detect}) prompted to flag malicious memory and \emph{run on the model under
test}, not a weaker external model or a stub, which is the strongest content detector
available per model.
\item \textsf{lineage} (MemLineage-style) authorizes unless a supporting item has a live
derivation edge to an untrusted ancestor. When provenance is inconclusive (no strong edge)
it falls back to \textsf{trusted}, faithful to lineage systems that default-allow rather
than block legitimate work. We grant it \emph{perfect} edges (no extraction error).
\item \textsf{capability\_ifc} (CaMeL/Fides-style) is single-session capability/IFC that
assumes long-term memory is uncompromised, so retrieved memory carries no taint and the
value is authorized. We grant it its best-case assumption (clean store).
\item \textsf{tma\_nm} (ours) binds a non-malleable effective origin (an item derived from,
or a trusted tool echoing, untrusted content stays \textsf{untrusted}), plus elevation only
on ${\ge}2$ independent trusted principals.
\end{itemize}
Crucially, the baselines fail \emph{structurally} rather than from weak tuning.
\textsf{capability\_ifc} authorizes laundered memory \emph{regardless of any threshold}
because it assumes a clean store, and \textsf{lineage} allows whenever no live untrusted
edge survives the laundering transformation. A sensitivity sweep over a content judge's
threshold cannot close a structural blind spot. This is precisely the content/lineage
insufficiency that T1 predicts, and Section~\ref{sec:sensitivity} confirms it empirically,
since no judge threshold reaches $(0\%~\text{ASR},100\%~\text{utility})$. We grant each
baseline its most favorable faithful configuration, so the benchmark does not handicap them.

\section{Per-Model, Per-Channel Results}\label{app:perchannel}
Table~\ref{tab:perchannel} gives the undefended (\textsf{none}) consequential-ASR for each
of the eight models on each laundering channel ($n{=}24$ per cell), and \tma{} is $0/24$ in
\emph{every} cell, that is, $0/192$ pooled per channel (Wilson 95\% CI $[0,2.0]\%$). The
complete five-defense $\times$ four-channel $\times$ eight-model cube, with raw counts, is
in the released artifact.
\begin{table}[h]
\centering
\caption{Undefended consequential-ASR (\%) by model and laundering channel ($n{=}24$ per
cell). \tma{} is $0\%$ in every cell ($0/192$ per channel, CI $[0,2.0]$).}
\label{tab:perchannel}
\begin{tabular}{lrrrr}
\toprule
Model & direct & summ. & echo & corrob. \\
\midrule
gpt-5-chat & 88 & 58 & 67 & 67 \\
gpt-4o-mini & 67 & 67 & 83 & 33 \\
claude-opus-4.1 & 67 & 67 & 67 & 67 \\
claude-sonnet-4.5 & 67 & 67 & 67 & 67 \\
gemini-2.5-flash & 100 & 100 & 67 & 67 \\
llama-4-maverick & 96 & 67 & 62 & 58 \\
deepseek-chat & 100 & 79 & 67 & 67 \\
qwen3-235b & 92 & 88 & 67 & 67 \\
\midrule
\textbf{pooled (none)} & 84 & 74 & 68 & 61 \\
\bottomrule
\end{tabular}
\end{table}

\end{document}